A tutorial overview on the angular scattering models of Electron-Neutral, Ion-Neutral, Neutral-Neutral, and Coulomb Collisions in Monte Carlo collision modeling on low temperature plasma

Wei Yang

Institute of Applied Physics and Computational Mathematics, Beijing, 100094, People's Republic of China

**Abstract** Over the past decade, massive modeling practices on low temperature plasma (LTP) reveal that input data such as microscopic scattering cross sections are crucial to output macroscopic phenomena. In Monte Carlo collision (MCC) modeling on LTP, angular scattering model is a non-trivial topic in both natural and laboratory plasma. Conforming to the pedagogical purpose of this overview, the classical and quantum theory of binary scattering including the commonly used Born-Bethe approximation is first introduced. State-of-the-art angular scattering models are derived based on the above theories for electron-neutral, ion-neutral, neutral-neutral, and Coulomb collisions. The tutorial is not aiming to provide accurate cross section data by modern approaches in quantum theory, but to introduce analytical angular scattering models from classical, semi-empirical, and first-order perturbation theory. The reviewed models are expected to be readily incorporated into the MCC codes, in which scattering angle is randomly sampled through analytical inversion instead of numerical accept-reject method. Those simplified approaches are very attractive, and demonstrate in many cases the ability to achieve a striking agreement with experiments. Energy partition models on electron-neutral ionization are also discussed with insight from the binary encounter Bethe theory. This overview is written in a tutorial style, in order to serve as a guide for novice in this field, and at the same time as a comprehensive reference for practitioners in MCC modeling on plasma.

# I. Introduction

Low temperature plasma (LTP) spans over many fields, such as advanced semiconductor processing, plasma medicine, electrical propulsion, combustion and hypersonic, agriculture and food cycle, etc. [1]. Numerical modeling is a basic method used to study LTP, to complement and/or assist experimental diagnostics [2]. Depending on the physical models of LTP, numerical modeling of different levels may be adopted: the kinetic methods based on direct solution to Boltzmann equation (BE), or statistical particle-in-cell (PIC) Monte Carlo (MC) collision approach, the fluid model based on velocity moments of BE, and the spatially-averaged global model. All the above models require input from microscopic parameters such as scattering cross sections or rate and transport coefficients. Therefore, different web-based databases are available [3-10], which collect, share, and verify cross sections, interacting potentials, as well as oscillator strength.

In Monte Carlo modeling of LTP, individual super-particle undergoes various types of collisions including elastic, excitation, ionization, charge-exchange, etc., in which the collision probability depends on the integrated cross section (ICS) and the deflection angle depends on the differential cross section (DCS) [2]. However, the DCS collected in different databases cannot be directly used in MC modeling, and the community strongly relies on isotropic scattering models or ill-mentioned forward peaking models. The first reason is due to lack of reasonably complete DCS data in the whole range of incident energies and scattering angles [11], which is limited by the resolution and range of the measuring system. The second reason is due to the lack of analytically simple-to-use models of DCS, through which scattering angles can be randomly sampled by easy-inversion instead of the time-consuming accept-reject method. During the last two decades, different angular scattering models have been compared including electron transport study, ionization source terms in glow discharge, thermal runaway electrons in streamer discharge, and interstellar neutral atoms in the outer heliosheath [12-27]. It is found that the angular scattering model is a non-trivial topic.

The purpose of this paper is to give a tutorial overview on angular scattering models in MC collision modeling on LTP, as well as energy partition models in electron-impact ionization. The tutorial is not aiming to provide accurate DCS data by modern approaches in quantum mechanics such as convergent close-coupling, R-matrix method, Schwinger multichannel approach, and local complex potential [28-32]. Instead, it is aiming to introduce analytical DCS models from classical, semi-empirical, and first-order perturbation theory in quantum mechanical approaches such as Born-Bethe approximation. Simplified approaches are very attractive in Monte Carlo collision modeling on LTP, because they demonstrate in many cases the ability to achieve a striking agreement with experiments. Conforming to the educational purpose of this tutorial, a brief overview of classical and quantum theory of DCS is introduced in Sec. II and III, respectively, and the latter includes first-order Born-Bethe approximation and partial wave analysis. State-of-the-art angular scattering models in MC modeling are introduced in Sec. IV in detail: neutral-neutral, charged particle Coulomb collision, ion-neutral, and electron-neutral elastic, excitation and ionization collision. Finally, a summary is given in Sec. V.

## II. Scattering cross sections in the classical theory - a brief overview

### a) Differential and integrated cross sections

The DCS of binary collision is schematically shown in Figure 1. The particle beam is incident at a fixed scattering center (If both projectile and target are moving, the coordinate system is transformed into the center of mass frame).

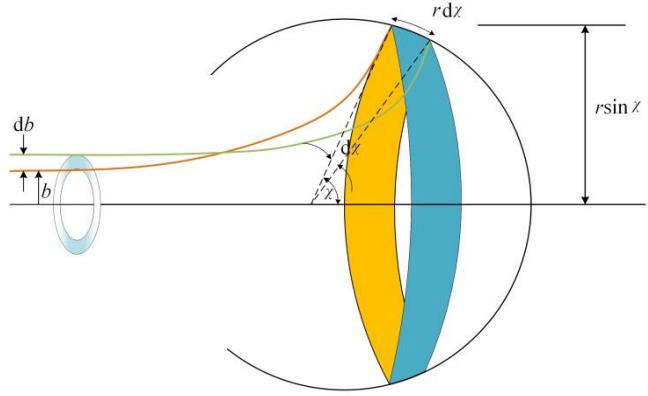

Fig. 1 The schematic of the differential scattering geometry.

The DCS $q(v, \chi)$ is the ratio between the number of particles deflected in the unit time in a given solid angle $dn/d\Omega$ and the incoming particle flux per unit time and per unit surface $j_{in}$.

$$q(v, \chi) = \frac{1}{j_{in}} \frac{dn}{d\Omega} \tag{1}$$

Assuming that the interaction force is symmetrical along the connecting line between the two particles, the incident particle with an impact parameter $b$ is scattered out in a determined direction of polar angle $\chi$, and the azimuth angle is unchanged. According to the conservation of particle, the scattered-out particle number is

$$dn = j_{in} 2\pi b db \tag{2}$$

In a binary collision, the azimuth angle $\phi$ is uniformly distributed between 0 and $2\pi$, and the differential solid angle

$$d\Omega = 2\pi \sin \chi d\chi \tag{3}$$

Inserting Eqs. (2) and (3) into the definition of DCS in Eq. (1), it is obtained that

$$q(v, \chi) = \frac{b}{\sin \chi} \left| \frac{db}{d\chi} \right| \tag{4}$$

The absolute value of $db/d\chi$ is used since its value is negative, namely a large impact parameter $b$ corresponds to a small scattering angle $\chi$. The total cross section $\sigma$ is the integral of the DCS over the whole solid angle

$$\sigma(v) = 2\pi \int_0^\pi W(\chi) q(v, \chi) \sin \chi \, d\chi \tag{5}$$

The weight function $W(\chi)$ is unity for total cross section $\sigma_t$, $1 - \sqrt{1 - E_{th}/E} \cos \chi$ for momentum transfer or diffusion cross section $\sigma_D$, and $1 - \sqrt{1 - E_{th}/E} \cos^2 \chi$ for energy transfer or viscosity cross section $\sigma_\mu$, where $E_{th}$ is the threshold energy for inelastic collision ($E_{th}=0$ for elastic collision).

The normalized DCS or the angular scattering function is defined as

$$I(v, \chi) = q(v, \chi) / \sigma_t(v) \tag{6}$$

which obeys the following normalization rule

$$2\pi \int_0^\pi I(v, \chi) \sin \chi \, d\chi = 1 \tag{7}$$

**b) The impact parameter and scattering angle**

The value of $db/d\chi$ is determined by the interacting potential. In a central-force potential, the motion of projectile relative to target can be reduced to the motion of particle with reduced mass $m_r = \frac{m_p m_t}{m_p + m_t}$ relative to a fixed center using the center-of-mass coordinate. Assuming that at a sufficient far distance the potential energy is zero, the initial relative velocity before collision is $v_0$, and the impact parameter is $b$, the conservation of angular momentum and energy reads

$$r^2 \frac{d\theta}{dt} = bv_0, \tag{8a}$$

$$\frac{1}{2} m_r (\frac{dr}{dt})^2 + \frac{1}{2} m_r r^2 (\frac{d\theta}{dt})^2 + V(r) = \frac{1}{2} m_r v_0^2. \tag{8b}$$

Combining the above equations and eliminating the time, the trajectory of the reduced particle is solved as follows

$$(\frac{dr}{d\theta})^2 = \frac{r^4}{b^2} - r^2 - \frac{2Vr^4}{m_r b^2 v_0^2} \tag{9}$$

Introducing a non-dimensional impact parameter as $W=b/r$, the trajectory is written in

a simple form

$$(\frac{dW}{d\theta})^2 = 1 - W^2 - \frac{V(r)}{E_r} \tag{10}$$

In Eq. (10), $E_r$ is the relative kinetic energy $E_r = m_r v_0^2 / 2$. Integrating Eq. (10), the relation between impact parameter and scattering angle is obtained.

$$\theta = \int_0^{b/r_{min}} (1 - W^2 - \frac{V(r)}{E_r})^{-1/2} dW \tag{11}$$

The limit $r_{min}$ is determined by equating the right hand side (RHS) of Eq. (10) to zero, which is simplified into

$$E_r \times (1 - \frac{b^2}{r_{min}^2}) - V(r_{min}) = 0 \tag{12}$$

Finally the scattering angle is obtained as

$$\chi = \pi - 2\theta = \pi - 2\int_0^{b/r_{min}} (1 - W^2 - \frac{V(r)}{E_r})^{-1/2} dW \tag{13}$$

### c) The model of inverse-power potential

Now a specific form of inverse-power potential $V(r) = Cr^{-n}$ is investigated, which in many cases describes the interacting potential of inter-particle scattering.

$$V(r) = Cr^{-n} \tag{14}$$

The potential energy to relative kinetic energy is written as

$$\frac{Cr^{-n}}{E_r} = (\frac{W}{W_0})^n, W_0 = b(\frac{E_r}{C})^{1/n} \tag{15}$$

According to the definition in Eq. (1), the DCS is derived

$$q d\Omega = W_0 (E_r / C)^{-2/n} dW_0 d\phi \tag{16}$$

The total cross section is infinite unless cutoff by a maximum impact parameter $W_{0M}$

$$\sigma_t = \int_0^{2\pi} \int_0^{W_{0M}} W_0 (E_r / C)^{-2/n} dW_0 d\phi = \pi W_{0M}^2 (E_r / C)^{-2/n} \tag{17}$$

However, the momentum and energy transfer cross section for elastic collision is finite and the cutoff of impact parameter is unnecessary.

$$\sigma_D = \int_0^{2\pi}\int_0^{\pi}(1-\cos\chi)q\sin\chi\,d\chi d\phi = 2\pi(E_r/C)^{-2/n}A_1(n+1) \tag{18a}$$

$$\sigma_\mu = \int_0^{2\pi}\int_0^{\pi}(1-\cos^2\chi)q\sin\chi\,d\chi d\phi = 2\pi(E_r/C)^{-2/n}A_2(n+1) \tag{18b}$$

The numerical constants $A_1$ and $A_2$ are calculated by the following integral.

$$A_1(n+1) = \int_0^{\infty}(1-\cos\chi)W_0 dW_0, \tag{19a}$$

$$A_2(n+1) = \int_0^{\infty}(1-\cos^2\chi)W_0 dW_0, \tag{19b}$$

There are two important conclusions in the case of inverse-power potential. Firstly, inserting Eq. (18b) into the Chapman-Enskog transport theory, a useful rule is drawn between the viscosity coefficient $\mu$ and the temperature as $\mu \propto T^{1/2+2/n}$.

$$\mu = \frac{5m(RT/\pi)^{1/2}(2mRT/C)^{2/n}}{8A_2(n+1)\Gamma(4-2/n)} \tag{20}$$

In Eq. (20), $\Gamma$ is the Gamma function.

Secondly, under the small-angle approximation ($\chi \ll 1$), the relation between impact parameter $b$ and scattering angle $\chi$ is explicitly expressed as

$$b = (\frac{A}{E_r\chi})^{1/n}, A = \frac{C\sqrt{\pi}\Gamma[(i+1)/2]}{\Gamma(i/2)} \tag{21}$$

Inserting Eq. (21) into the expression of DCS Eq. (4),

$$q(v,\chi) = \frac{1}{n}(\frac{A}{E_r})^{2/n}\frac{1}{\chi^{(2/n+1)}\sin\chi} \tag{22}$$

Under the small-angle scattering $\sin\chi \approx \chi$, the DCS is expressed as the inverse-power of both relative kinetic energy and scattering angle

$$q(v,\chi) = \frac{1}{n}(\frac{A}{E_r})^{2/n}\frac{1}{\chi^{(2/n+2)}} \tag{23}$$

## III. Scattering cross sections in the quantum theory

### a) Cross section and the scattering amplitude

In quantum mechanics, there is a probability that the projectile changes its momentum due to scattering with the target center. The DCS is related to this probability according to the definition in Eq. (1), the evaluation of which also requires $j_{in}$ and $dn/d\Omega$. The particle flux $j$ is calculated by the wave function as

$$j = \frac{\hbar}{2im_r}(\varphi^* \nabla \varphi - \varphi \nabla \varphi^*) \tag{24}$$

In the simplified approach, plane wave function is used for freely incident particle

$$\varphi_{inc} = e^{i\mathbf{k}\cdot\mathbf{r}} \tag{25}$$

Inserting Eq. (25) into Eq. (24), the incident particle flux is obtained

$$j_{inc} = \frac{\hbar k}{m_r} \tag{26}$$

Assuming that the interaction force is symmetrical along the connecting line between the two particles, the wave-function demonstrates the following asymptotic behavior when radial position approaches infinite $r \to \infty$

$$\varphi = \varphi_{inc} + \varphi_{sca} = e^{i\mathbf{k}\cdot\mathbf{r}} + f(\mathbf{k} \to \mathbf{k}_{sca})\frac{e^{ik_{sca}r}}{r} \tag{27}$$

In Eq. (27), $f(\mathbf{k} \to \mathbf{k}_{sca})$ is the scattering amplitude. Inserting Eq. (27) into Eq. (24), the scattered-out particle flux is obtained

$$j_{out} = \frac{\hbar k_{sca}}{m_r}\frac{1}{r^2}|f(\mathbf{k} \to \mathbf{k}_{sca})|^2 \tag{28}$$

The number of particles scattered out in the unit time in a given solid angle $dn/d\Omega$

$$\frac{dn}{d\Omega} = \frac{\hbar k_{sca}}{m_r}|f(\mathbf{k} \to \mathbf{k}_{sca})|^2 \tag{29}$$

Inserting Eq. (26) and (29) into Eq. (1), the DCS in quantum theory is derived

$$q(v,\chi) = \frac{k_{sca}}{k}|f(\mathbf{k} \to \mathbf{k}_{sca})|^2 \tag{30}$$

Once the scattering amplitude f($\mathbf{k} \to \mathbf{k_{sca}}$) is obtained, the DCS is readily determined. A variety of modern methods, such as R-matrix, convergent close-coupling, Schwinger multichannel approach, local complex potential are used in accurate calculation of DCS [28-32]. However, those models are beyond the scope of this paper, because they typically do not provide an analytical DCS for easy-inversion. Instead, some approximation methods, including Born-Bethe approximation and partial wave analysis are commonly used to fit the experimental DCS in MC collision modeling.

### b) First-order Born-Bethe approximation

In the case of high energy collision, Born approximation is one of the simplest ways to evaluate the cross section. It is not necessary to solve the Schrodinger's equation in differential form,

$$(\nabla^2 + k^2)\varphi(\mathbf{r}) = \frac{2m_r}{\hbar^2} V(\mathbf{r})\varphi(\mathbf{r}) \tag{31}$$

but the Lippman-Schwinger equation in integral form.

$$\varphi(\mathbf{r}) = e^{i\mathbf{k}\cdot\mathbf{r}} + \frac{2m_r}{\hbar^2} \int d^3r' G(\mathbf{r},\mathbf{r}')V(\mathbf{r}')\varphi(\mathbf{r}') \tag{32}$$

In Eq. (32), the first and second term in the RHS denote the incident plane wave, and the scattered wave respectively, and $G(\mathbf{r},\mathbf{r}')$ is the Green's function which satisfies

$$(\nabla^2 + k^2)G(\mathbf{r},\mathbf{r}') = \delta(\mathbf{r}-\mathbf{r}'), \tag{33}$$

and has the following properties

$$G(\mathbf{r},\mathbf{r}') = G(\mathbf{r}-\mathbf{r}') = -\frac{\exp(ik|\mathbf{r}-\mathbf{r}'|)}{4\pi|\mathbf{r}-\mathbf{r}'|} \tag{34}$$

Inserting Eq. (34) into Eq. (32), the wave function is written as

$$\varphi(\mathbf{r}) = e^{i\mathbf{k}\cdot\mathbf{r}} - \frac{m_r}{2\pi\hbar^2} \int d^3r' \frac{\exp(ik|\mathbf{r}-\mathbf{r}'|)}{|\mathbf{r}-\mathbf{r}'|} V(\mathbf{r}')\varphi(\mathbf{r}') \tag{35}$$

When evaluating the integral in the second term, the wave function can be replaced by the incident plane wave in a first-order Born approximation (FBA).

$$\varphi(\mathbf{r}) \approx e^{i\mathbf{k}\cdot\mathbf{r}} - \frac{\mu}{2\pi\hbar^2}\int d^3r' \frac{\exp(ik|\mathbf{r}-\mathbf{r}'|)}{|\mathbf{r}-\mathbf{r}'|}V(\mathbf{r}')e^{i\mathbf{k}\cdot\mathbf{r}'} \tag{36}$$

Based on the asymptotic behavior of the scattered wave function at infinite radial position r → ∞, the scattering amplitude is obtained.

$$f(\mathbf{k} \to \mathbf{k}_{sca}) = -\frac{m_r}{2\pi\hbar^2}\int d^3r \exp(i\mathbf{K}\cdot\mathbf{r})V(\mathbf{r}), \mathbf{K} = \mathbf{k} - \mathbf{k}_{sca} \tag{37}$$

Finally the DCS in FBA is derived [33]

$$q^{Born} = \frac{k_{sca}}{k}|f(\mathbf{k}\to\mathbf{k}_{sca})|^2 = \frac{m_r^2}{4\pi^2\hbar^4}\frac{k_{sca}}{k}\left|\int d^3r \exp(i\mathbf{K}\cdot\mathbf{r})V(\mathbf{r})\right|^2 \tag{38}$$

In many cases, the incident wave function is not represented by the plane wave, but the eigen-functions $u_i$ of the system

$$q^{Born} = \frac{m_r^2}{4\pi^2\hbar^4}\frac{k_{sca}}{k}\left|\int d^3r' \exp(i\mathbf{K}\cdot\mathbf{r}')u_n^*V(\mathbf{r}')u_0\right|^2 \tag{39}$$

Bethe applied the Born approximation to a specific process in which a particle of charge $ze$ collides with an atom [34] with the following interacting potential

$$V(r) = \frac{zZ_N e^2}{r} - \sum_{j=1}^{Z}\frac{ze^2}{|\mathbf{r}-\mathbf{r}_j|} \tag{40}$$

In Eq. (40), $Z_N$ is the charge of the atomic nucleus and $\mathbf{r}_j$ the coordinate of the atomic electrons. Bethe recognized that it is more convenient to perform integration over $\mathbf{r}$ by the relation

$$\int d\mathbf{r}\frac{\exp(i\mathbf{K}\cdot\mathbf{r})}{|\mathbf{r}-\mathbf{r}_j|} = \frac{4\pi}{K^2}\exp(i\mathbf{K}\cdot\mathbf{r}_j) \tag{41}$$

Thus, Eq. (39) transforms into

$$q^{Bethe} = \frac{4m_r^2 z^2 e^4}{\hbar^4 K^4}\frac{k_{sca}}{k}|\varepsilon_n(\mathbf{K})|^2 \tag{42}$$

In Eq. (42), $\varepsilon_n(\mathbf{K})$ is called the matrix element

$$\varepsilon_n(\mathbf{K}) = \langle n|\sum_{j=1}^{Z}\exp(i\mathbf{K}\cdot\mathbf{r}_j)|0\rangle \tag{43}$$

Under many situations, the matrix element is a function of scalar variable $K$, which is more convenient than scattering angle $\chi$.

$$K^2 = k^2 + k_{sca}^2 - 2kk_{sca}\cos\chi \tag{44}$$

Therefore, one would replace the differential solid angle by

$$d\Omega = 2\pi \sin\chi d\chi = 2\pi \frac{K}{kk_{sca}} dK \tag{45}$$

to obtain

$$d\sigma = \frac{8\pi m_r^2 z^2 e^4}{\hbar^4 K^3 k^2} |\varepsilon_n(K)|^2 dK \tag{46}$$

Also, one could use the dimension of energy $Q = \frac{\hbar^2 K^2}{2m_r}$ and write

$$d\sigma = \frac{\pi z^2 e^4}{Q E_r} |\varepsilon_n(K)|^2 d(\ln Q) \tag{47}$$

In atomic physics, the general oscillator strength (GOS) is commonly used instead of the matrix element

$$f_n(K) = |\varepsilon_n(K)|^2 \frac{E_n}{Q} \tag{48}$$

When the momentum change approaches zero $K \to 0$, the optical oscillator strength (OOS) $f_n$ is derived

$$f_n = \frac{E_n}{E_{Ryd}} |M_n|^2, \quad |M_n|^2 = \frac{|\int u_n^* \Sigma_{j=1}^Z x_j u_0 d\mathbf{r}_1 \dots d\mathbf{r}_Z|^2}{a_0^2} \tag{49}$$

In Eq. (49), $E_{Ryd}$ is Rydberg constant and $a_0$ is the Bohr radius. The OOS is proportional to the optical absorption cross section, whose value is well documented in several databases. Based on the GOS, the DCS of FBA can be evaluated. Born approximation yields an order-of-accurate cross section if neither experimental nor theoretical data are available. If appropriate scaling law is used [35], the accuracy of the scaled Born cross section is comparable to that of the experiment.

### c) Partial-wave analysis

In the case of low energy collision, partial wave analysis is a strict method if sufficient numbers of wave function is used. Scattering from a spherical potential, the orbital wave function is expanded in terms of spherical harmonic function (the spin of particle is neglected)

$$\varphi = \sum_{l}^{\infty} R_l(kr)\, Y_{l0}(\chi) \tag{50}$$

Inserting Eq. (50) into the Schrödinger equation (31), partial wave with different $l$ in radial direction is decomposed

$$\left[\frac{1}{r^2}\frac{d}{dr}r^2\frac{d}{dr} + k^2 - \frac{l(l+1)}{r^2} - \frac{2m_r V}{\hbar^2}\right] R_l = 0 \tag{51}$$

In Eq. (50), the radial wave function can be written as a sum of incident wave and outward going wave

$$R_l(kr) = \sqrt{4\pi(2l+1)}\, i^l [j_l(kr) + \frac{a_l}{2} h_l(kr)] \tag{52}$$

In Eq. (52), $j_l$ and $h_l$ is the spherical Bessel function of the first and second kind, respectively. Now considering the asymptotic behavior of the scattered wave function at infinite radial position $r \to \infty$, the scattering amplitude is obtained as

$$f = \frac{1}{2ik}\sum_{l=0}^{\infty}(2l+1)(S_l - 1)P_l(\cos\chi),\, S_l = 1 + a_l \tag{53}$$

Elastic collision requires that the amplitude of partial wave with any $l$ is unchanged $|S_l| = 1$, therefore,

$$1 + a_l = e^{2i\eta_l} \tag{54}$$

In Eq. (54), $\eta_l$ is called phase shift. The partial wave function of $l$ at infinite radial position is related to corresponding phase shift

$$\varphi_l = \frac{2l+1}{k} e^{i\eta_l} \sin(\eta_l)\, P_l(\cos\chi)\, \frac{e^{ikr}}{r} \tag{55}$$

The scattering amplitude is the sum of contributions from all partial waves

$$f_\chi = \sum_{l=0}^{\infty}\frac{2l+1}{k} e^{i\eta_l} \sin(\eta_l)\, P_l(\cos\chi) = \frac{1}{2ik}\sum_{l=0}^{\infty}(2l+1)(e^{2i\eta_l} - 1)P_l(\cos\chi) \tag{56}$$

Finally, the DCS is derived

$$q = |\sum_{l=0}^{\infty}\frac{2l+1}{k} e^{i\eta_l} \sin(\eta_l)\, P_l(\cos\chi)|^2 = \frac{4\pi}{k^2}|\sum_{l=0}^{\infty}\sqrt{2l+1}\, e^{i\eta_l} \sin(\eta_l)\, Y_{l0}(\chi)|^2 \tag{57}$$

Therefore, the evaluation of elastic DCS requires the information of scattering amplitude of each partial wave, which only depends on the phase shift solved in Eq. (51). In practice, the number of partial waves is cutoff depending on the relative kinetic energy. There are two simple methods including the Born approximation and effective range theory to determine the phase shift.

Considering a high-energy particle scattered from a spherical potential, the scattering amplitude in FBA Eq. (38) is integrated in spherical coordinate

$$f(\chi) = f(\mathbf{k} \to \mathbf{k}_{sca}) = -\frac{2\mu}{\hbar^2 K}\int_0^\infty r\sin(Kr)V(r)\,\mathrm{d}r \tag{58}$$

By the following relation

$$\frac{\sin(Kr)}{Kr} = \sum_{l=0}^{\infty}(2l+1)j_l^2(kr)P_l(\cos\theta), \tag{59}$$

The phase shift is solved

$$e^{i\eta_l}\sin\eta_l = -\frac{2\mu k}{\hbar^2}\int_0^\infty V(r)\,j_l^2(kr)r^2\mathrm{d}r \tag{60}$$

Effective range theory has its root in scattering studies involving nuclear interacting potential. The potential energies are expressed as functions of the well depth and the intrinsic range beyond which the potential is zero. The intrinsic range is defined by two parameters, namely effective range and scattering length, hence the name effective range theory is called. Modified effective range theory (MERT) is used in electron collision with polarizable atoms and molecules [36], which utilizes the phase shift representation of quantum-mechanical DCS in low-energy collision up to a few electron Volts. For partial wave with different number $l$, the commonly used analytical phase shifts to fit the experimental obtained cross section are as follows.

$$\tan\eta_0 = -Ak\left[1 + \frac{4\alpha_p}{3a_0}k^2\ln(3a_0)\right] - \frac{\pi\alpha_p}{3a_0}k^2 + Dk^3 + Fk^4 \tag{61a}$$

$$\tan\eta_1 = \frac{\pi\alpha_p}{15a_0}k^2[1 - \sqrt{E/E_1}] \tag{61b}$$

$$\tan\eta_l = \frac{\pi\alpha_p}{(2l+3)(2l+1)(2l-1)a_0}k^2 \tag{61c}$$

In Eq. (61a-c), A, D, F, $E_1$ are fitting parameters, $\alpha_p$ is the polarizability.

### d) Scattering of identical particles

The scattering of two identical particles lead to the exchange effect in quantum mechanics. The orbital wave function should be either symmetric or asymmetric [37], and the asymptotic behavior in Eq. (27) is written as

$$\varphi = e^{i\mathbf{k}\cdot\mathbf{r}} \pm e^{-i\mathbf{k}\cdot\mathbf{r}} + \frac{e^{ik_{sca}r}}{r}[f(\chi) \pm f(\pi-\chi)] \tag{62}$$

The symmetric and asymmetric DCS is written as

$$q_{s,a} = |f(\chi) \pm f(\pi-\chi)|^2 = |f(\chi)|^2 + |f(\pi-\chi)|^2 \pm 2\,\text{Re}[f(\chi)f^*(\pi-\chi)] \quad (63)$$

The last term is due to the interference effect of quantum-mechanical wave function. If the interference term is not included, Eq. (63) reduces to the classical DCS

$$q_{\text{total}} = |f(\chi)|^2 + |f(\pi-\chi)|^2 = q(\chi) + q(\pi-\chi) \quad (64)$$

Let us now consider the particle with spin. If the total spin of the two identical particles is not fixed, the DCS is averaged over all possible spin states, assuming each spin state has an equal probability. If the spin of identical particle $s$ is half integer, the total DCS is

$$q_{\text{total}} = \frac{s}{2s+1}q_s + \frac{s+1}{2s+1}q_a = |f(\chi)|^2 + |f(\pi-\chi)|^2 - \frac{2}{2s+1}\text{Re}[f(\chi)f^*(\pi-\chi)] \quad (65)$$

If the spin of identical particle $s$ is integer, the total DCS is

$$q_{\text{total}} = \frac{s+1}{2s+1}q_s + \frac{s}{2s+1}q_a = |f(\chi)|^2 + |f(\pi-\chi)|^2 + \frac{2}{2s+1}\text{Re}[f(\chi)f^*(\pi-\chi)] \quad (66)$$

## IV. Angular scattering models in Monte Carlo collision modeling

In MC collision modeling on binary scattering, the azimuth angle is randomly sampled from a uniform distribution between 0 and $2\pi$, and the polar angle $\chi$ is randomly sampled according to the angular scattering model

$$2\pi \int_0^\chi I(v,\chi')\sin\chi'\,\mathrm{d}\chi' = r \quad (67)$$

In Eq. (67), $r$ is a random number uniformly distributed between [0, 1]. If the angular scattering model is analytical and easy to invert, the sampling process is computationally efficient. Otherwise, time-consuming sampling methods may be adopted such as accept-reject method. Note that in some cases, random sampling of the classical impact parameter $b$ or quantum-mechanical wave-vector change $K$ is more convenient than Eq. (67), then scattering angle is derived by the sampled $b$ or $K$ (see Appendix a for examples).

The following part of this section will overview state-of-the-art angular scattering models based on classical and quantum theory, and also empirical models which are commonly used in MC collision modeling. The models reviewed in this section are not claimed to be complete, rather they should be representative and reasonable.

**a) Neutral-neutral collision: variable hard and soft sphere**

Neutral-neutral collision is well described by the classical theory. According to hard sphere model, the impact parameter is determined by the diameter of the sphere and the scattering angle as schematically shown in figure 2

$$b = (a_1 + a_2)\cos\frac{\chi}{2} = d\cos\frac{\chi}{2} \tag{68}$$

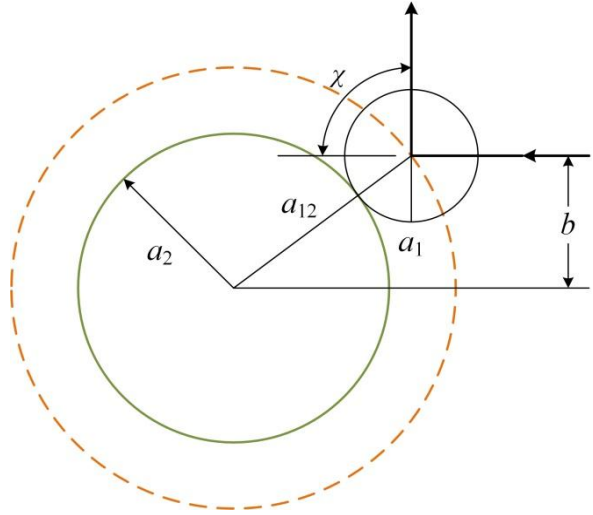

Fig. 2 The schematic of the hard sphere model.

The DCS is obtained according to Eq. (4)

$$q(v,\chi) = \frac{1}{4}(a_1 + a_2)^2 \tag{69}$$

The DCS is a constant, irrespective of the incident energy and scattering angle. Inserting $I(v,\chi) = \frac{1}{4\pi}$ into Eq. (67), the scattering angle is sampled as

$$\cos\chi = 1 - 2r \tag{70}$$

The DCS of hard sphere model is representative of isotropic scattering, which yields the following relation between transport cross sections

$$\sigma_D / \sigma_t = 1, \sigma_\mu / \sigma_t = 2/3 \tag{71}$$

In fact, the hard sphere model can be viewed as a case of $n=\infty$ in inverse-power potential. The viscosity of hard sphere model is obtained as

$$\mu = \frac{5}{16} \frac{(\pi mkT)^{1/2}}{\sigma_T} = \frac{5}{16} \frac{(mkT/\pi)^{1/2}}{d^2} \tag{72}$$

Eq. (72) can be used to determine the diameter of hard sphere $d$ if the viscosity at a given temperature is known.

The hard sphere model, however, is too simple to be used for real gases. Therefore, variable hard sphere (VHS) model is proposed by Bird [38], which assumes the cross section as an inverse power of relative kinetic energy

$$\frac{\sigma_t}{\sigma_{ref}} = (\frac{d}{d_{ref}})^2 = (\frac{E}{E_{ref}})^{-\xi} = (\frac{v_r}{v_{ref}})^{-2\xi} \tag{73}$$

Comparing Eq. (73) with the cross section of inverse-power potential Eq. (17), it is determined that $\xi = 2/n$. The angular scattering model of VHS still inherits the isotropic character of hard sphere model, and the scattering angle is sampled according to Eq. (70), just as the case $n=\infty$. One should note that the ICS and DCS of VHS model adopts a different power exponent in the inverse-power potential!

The transport cross section of VHS can be evaluated by Eq. (71), but departs significantly from measurements in some gases. Koura and Matsumoto further proposed the variable soft sphere (VSS) model [39, 40], in which a soft parameter $\alpha$ is introduced in the relation between the impact diameter and the diameter of the sphere

$$b = d \cos^\alpha \frac{\chi}{2} \tag{74}$$

Inserting Eq. (74) into Eq. (4), the DCS of VSS is determined as

$$q(v, \chi) = \frac{1}{4} d^2 \cos^{2\alpha-2} \frac{\chi}{2} \tag{75}$$

The transport cross sections is readily derived by integration

$$\sigma_t = \pi d^2, \sigma_m = \frac{2}{\alpha+1}\sigma_t, \sigma_\mu = \frac{4\alpha}{(\alpha+1)(\alpha+2)}\sigma_t \tag{76}$$

The value of $\alpha$ is determined according to Eq. (18)

$$\frac{\sigma_\mu}{\sigma_m} = \frac{2\alpha}{\alpha+2} = \frac{A_2(n+1)}{A_1(n+1)} \tag{77}$$

Therefore the soft parameter $\alpha$ is obtained as

$$\alpha = [\frac{A_1(n+1)}{A_2(n+1)} - \frac{1}{2}]^{-1} \tag{78}$$

The diameter of the soft/hard sphere is calculated according to the viscosity at a reference temperature

$$d = \left(\frac{5(\alpha+1)(\alpha+2)(m/\pi)^{0.5}(kT_{ref})^{0.5+\xi}}{16\alpha\Gamma(4-\xi)\mu_{ref}E_r^\xi}\right)^{0.5} \tag{79}$$

The literature [40, 41] usually documents the parameter $\alpha$, $\xi$, and the diameter at a reference temperature $d_{ref}$

$$d = d_{ref}[\frac{(kT_{ref}/E_r)^\xi}{\Gamma(2-\xi)}]^{0.5}, d_{ref} = \left(\frac{5(\alpha+1)(\alpha+2)(mkT_{ref}/\pi)^{0.5}}{16\alpha(3-\xi)(2-\xi)\mu_{ref}}\right)^{0.5} \tag{80}$$

Inserting the angular scattering function $I(v,\chi) = \frac{1}{4\pi}\cos^{2\alpha-2}\frac{\chi}{2}$ into Eq. (67), the scattering angle is randomly sampled as

$$\cos\chi = 2(1-r)^{1/\alpha} - 1 \tag{81}$$

There is another method to derive the sampling of the scattering angle as shown in Appendix a.

**b) Electron-electron, electron-ion, and ion-ion Coulomb collision**

The Coulomb collision between charged particle is a special case of inverse-power potential.

$$C = \frac{q_1 q_2}{4\pi\varepsilon_0}, n = 1 \tag{82}$$

Following common procedure in Sec. II (b) and (c), Rutherford first derived the DCS in classical theory for alpha particle scattering. The impact parameter is

$$b = \frac{b_0}{2\tan\frac{\chi}{2}},$$ (83)

In Eq. (83), $b_0$ is the classical distance of closest approach $b_0 = \frac{q_1 q_2}{4\pi\varepsilon_0 E_r}$. The DCS is obtained by inserting Eq. (83) into Eq. (4)

$$q(v,\chi) = \left(\frac{b_0}{4\sin^2\frac{\chi}{2}}\right)^2 = \frac{b_0^2}{4}\frac{1}{(1-\cos\chi)^2}$$ (84)

In quantum mechanical theory, the same DCS is also derived later by Wentzel [42] and Oppenheimer [43] using FBA. Shortly after, Mott obtained the following scattering amplitude using an accurate method [44]

$$f(\chi) = -\frac{1}{2\rho^2 \sin^2\frac{\chi}{2}}\frac{\Gamma(1+i/\rho)}{\Gamma(1-i/\rho)}\exp\left(-\frac{2i}{\rho}\ln\sin\frac{\chi}{2}\right), \rho^2 = \frac{pv}{q_1 q_2/(4\pi\varepsilon_0)} = \frac{2}{b_0}$$ (85)

The DCS calculated by Eq. (85) is the same as that of classical Rutherford's DCS. For electron-electron collision, the identical-particle effect in quantum mechanics can be included by inserting $s=1/2$ into Eq. (65). The DCS is obtained as

$$q_{e-e} = |f(\chi)|^2 + |f(\pi-\chi)|^2 - \text{Re}[f(\chi)f^*(\pi-\chi)]$$ (86)

Inserting the scattering amplitude Eq. (85) into Eq. (86), the DCS is written as

$$q(v,\chi) = \left(\frac{1}{2\rho_{ee}^2}\right)^2\left[\frac{1}{\sin^4\frac{\chi}{2}} + \frac{1}{\cos^4\frac{\chi}{2}} - \frac{1}{\sin^2\frac{\chi}{2}\cos^2\frac{\chi}{2}}\cos\left(\frac{1}{\rho_{ee}}\ln\tan^2\frac{\chi}{2}\right)\right],$$

$$\rho_{ee}^2 = \frac{m_{ee}v^2}{e^2/(4\pi\varepsilon_0)}$$ (87)

The above DCS is derived in the center-of-mass (COM) frame, and it can be transformed to the one-particle-at-rest frame by simply replace χ as 2ϑ (in general cases, readers are referred to the relation of scattering angle between COM and laboratory frame in classical textbooks)

$$q(v,\vartheta) = 4\left(\frac{1}{2\rho_{e-e}^2}\right)^2\left[\frac{1}{\sin^4\vartheta} + \frac{1}{\cos^4\vartheta} - \frac{1}{\sin^2\vartheta\cos^2\vartheta}\cos\left(\frac{1}{\rho_{e-e}}\ln\tan^2\vartheta\right)\right]\cos\vartheta$$ (88)

Noting that the differential solid angle is

$$\sin \chi \, d\chi \, d\varphi = 4 \sin \vartheta \cos \vartheta \, d\vartheta \, d\varphi \tag{89}$$

The energy of scattered electron is related to the scattering angle as

$$W = E \sin^2 \vartheta, E - W = E \cos^2 \vartheta \tag{90}$$

and the differential solid angle is also related to the differential energy range by

$$\cos \vartheta d\Omega = 2\pi \sin \vartheta \cos \vartheta \, d\vartheta = \frac{\pi dW}{E} \tag{91}$$

Inserting the above Eqs. (89-91) into Eq. (88), the DCS for energy is obtained as

$$\frac{d\sigma}{dW} = (\frac{1}{2\rho^2})^2 \frac{4\pi}{E} \{\frac{1}{W^2} + \frac{1}{(E-W)^2} - \frac{1}{W(E-W)} \cos(\sqrt{\frac{E_H}{E}} \ln[\frac{E-W}{E}])\} \tag{92}$$

If one of the electron energies $W$ or $E-W$ is much smaller than the incident energy $E$, the above DCS reduces to the classical Rutherford's one.

### i. TA's method

Takizuka and Abe (TA) proposed one of the earliest binary algorithms for MCC modeling on Coulomb scattering [45]. In TA's model, the particles are paired locally in space and undergo binary elastic scattering events. Under small-angle approximation, the average deflecting angle of one collision is determined by

$$\left\langle \frac{\chi^2}{4} \right\rangle_1 \approx \left\langle \tan^2 \frac{\chi}{2} \right\rangle = \frac{1}{\pi b_{max}^2} \int_{b_{min}}^{b_{max}} (\frac{b_0}{2b})^2 2\pi b \, db = \frac{1}{2}(\frac{b_0}{b_{max}})^2 \ln b \Big|_{b_{min}}^{b_{max}} \tag{93}$$

To obtain finite scattering angle, the minimum and maximum of impact parameter is cutoff

$$b_{max} = \lambda_D, b_{min} = b_0 / 2 . \tag{94}$$

Inserting Eq. (94) into Eq (93), the expectation of squared angle is derived

$$\left\langle \chi^2 \right\rangle_1 = \frac{2b_0^2}{b_{max}^2} \ln \Lambda, \Lambda = \frac{2\lambda_D}{b_0} \tag{95}$$

In Eq. (95), $\ln \Lambda$ is the Coulomb logarithm. In time step $\Delta t$, the expectation of squared angle is

$$\left\langle \chi^2 \right\rangle (\Delta t) = \left\langle \chi^2 \right\rangle_1 n(\pi b_{\max}^2) v_r \Delta t = 2\pi b_0^2 n v_r \Delta t \ln \Lambda \tag{96}$$

According to the central limit theorem, the variable $\delta = \tan \frac{\chi}{2}$ obeys the Gaussian distribution such that the average value is zero and variance is given by

$$\left\langle \delta^2 \right\rangle = \frac{1}{2} \pi b_0^2 n v_r \Delta t \ln \Lambda = \frac{1}{2} \left( \frac{q_1 q_2}{4\pi \varepsilon_0 E_r} \right)^2 \pi n v_r \Delta t \ln \Lambda \tag{97}$$

The time-step constraint for TA's method is limited by

$$\left\langle \delta^2 \right\rangle = v_{coll} \Delta t \ll 1, v_{coll} = \frac{(q_1 q_2)^2 n \ln \Lambda}{8\pi^2 \varepsilon_0^2 m_r^2 v_r^3} \tag{98}$$

Wang *et al.* [46] improved TA's method, pointing out the relation between the proposed collision operator and the Landau-Fokker-Planck (L-F-K) operator. The scattering angle is randomly sampled with a small parameter $\tilde{\varepsilon}$, which ensures that the friction coefficient and diffusion tensor is correctly reproduced in the limit of infinitesimally small time step.

$$\cos \chi = 1 - (1 - \cos \tilde{\varepsilon}) \sin^2 \theta, \tilde{\varepsilon}^2 = \frac{3(q_1 q_2)^2 n \ln \Lambda}{4\pi \varepsilon_0^2 m_r^2 v_r^3}, \cos \theta = 1 - 2r \tag{99}$$

Note that if the particle density $n_1$ and $n_2$ is not equal, or the particle number $N$ is not even, the choice of $n$ in the definition of $\tilde{\varepsilon}$ is slightly different.

### ii. Nanbu's method

However, simulating small-angle collisions one by one is computationally inefficient. Nanbu grouped a succession of small-angle binary collisions into a unique binary collision with a cumulative scattering angle $\chi_N$

$$\left\langle \sin^2 \frac{\chi_N}{2} \right\rangle = \frac{1}{2}(1 - \left\langle \cos \chi_N \right\rangle) = \frac{1}{2}(1 - e^{-s}) \tag{100}$$

The variable $s$ in Eq. (100) was proposed as a function of squared angle

$$s = \frac{1}{2} \left\langle \chi^2 \right\rangle (\Delta t) = 2 v_{coll} \Delta t \tag{101}$$

Through MC simulation of small-angle scattering [47], the probability distribution function (PDF) of $\chi_N$ was found to obey the following relation

$$\ln f(\chi_N) \propto \cos \chi_N \tag{102}$$

The normalized PDF is written as (actually a Gaussian distribution similar to TA's method)

$$f(\chi_N) = \frac{A}{4\pi \sinh A}\exp(A\cos \chi_N) = \frac{A}{2\pi(1-e^{-2A})}\exp(-2A\sin^2 \frac{\chi_N}{2}), \tag{103}$$

with the normalized parameter $A$ given by

$$\coth A - A^{-1} = e^{-s} \tag{104}$$

The transcendental Eq. (104) can be solved numerically according to the range of $s$, and look-up tables [47] or interpolation fitting [48] can be used. Once the parameter $A$ is obtained, the scattering angle is randomly sampled as

$$\cos \chi_N = \frac{1}{A}\ln(e^{-A} + 2r\sinh A) \tag{105}$$

Numerical convergence test showed that for both ion-ion and electron-ion collisions, the point-wise errors for the Nanbu's method running at approximately half the time-step were comparable to those of the TA's method [49].

In order to accelerate the MC scheme on Coulomb collision, a hybrid scheme was developed by Caflisch et al. [50], representing the velocity distribution function as a combination of a thermal component and a kinetic component. Dimits *et al.* [51] showed that Nanbu's model can be derived from an approximation to the Coulomb-Lorentz (C-L) kernel by a combination of Legendre polynomial

$$f_{CL}(\cos \chi) = \frac{1}{2\pi}\sum_{l=0}^{\infty}(l+\frac{1}{2})P_l(\cos \chi)\exp(-l(l+1)s) \tag{106}$$

Therefore, a more general PDF was proposed by composing Nanbu's distribution and the cutoff C-L kernel

$$f(\cos \chi) = \begin{cases} \dfrac{A}{4\pi \sinh A}\exp(A\cos \chi_N), s \leq s_0 \\ \dfrac{1}{2\pi}\sum_{l=0}^{m(s)}(l+\dfrac{1}{2})P_l(\cos \chi)\exp(-l(l+1)s), s > s_0 \end{cases} \tag{107}$$

In Eq. (107), $m(s)$ is a cutoff number of the sum. The Cumulative Distribution Function (CDF) was also calculated in order to randomly sample the scattering angle

$$F(\cos \chi) = 2\pi \int_{\cos \chi}^{-1} f(\cos \chi') \sin \chi' \mathrm{d}\chi' \tag{108}$$

Inserting Eq. (107) into Eq. (108), it gives

$$F = \begin{cases} \dfrac{A}{2 \sinh A}[\exp(A\cos \chi) - \exp(-A)], s \leq s_0 \\ \dfrac{1}{2} \sum_{l=0}^{m(s)} \{[P_{l+1}(\cos \chi) - P_{l-1}(\cos \chi)] - [P_{l+1}(-1) - P_{l-1}(-1)]\} \exp(-l(l+1)s), s > s_0 \end{cases} \tag{109}$$

Nanbu's method is capable to model the cumulative scattering of small angle, but in practice cannot capture the large angle scattering in single event. Recently, methods of combining the cumulative scattering events by a Gaussian-like PDF and a single-scattering event by the original Rutherford's DCS is proposed [52, 53].

### iii. DSMC-like method

In both TA's and Nanbu's method, all charged particles in the same spatial cell are paired together for MC collision. Is it possible that only portion of them are paired up instead of the whole $N_{\mathrm{coll}} = N(N-1)/2$? The general approach to DSMC methods for the Boltzmann equation with long range potentials (infinite total cross-section) and for the L-F-K equation was proposed by Bobylev and Nanbu [54]. Their work also showed that Nanbu's original model was really a method of solving the L-F-K equation for limiting case of small-angle scattering.

Shortly after, Dimarco *et al.* [55], Bobylev and Potapenko [56] developed the DSMC-like method, in which different scattering laws were proposed. Here, the simpler scattering law by Bobylev and Potapenko [56] is introduced, based on an approximation of L-F-K equation by Boltzmann equations of **quasi-Maxwellian** kind. The approach was claimed to include the TA's and Nanbu's original method as particular cases. The total number of collision pair in multi-component plasma is

$$N_{\mathrm{coll}} = \dfrac{N\nu \Delta t}{\tilde{\varepsilon}}, \nu = \dfrac{n \ln \Lambda}{8\pi \varepsilon_0^2} \max(\dfrac{q_i^2 q_j^2}{m_{ij}^2 v_0^3}) \tag{110}$$

In Eq. (110), $\tilde{\varepsilon}$ is a dimensionless parameter, $m_{ij}$ is the reduced mass of the colliding pair, and $v_0$ is a characteristic velocity which in practice can be the thermal velocity. The colliding pair is chosen in two steps. Firstly, the species of the collision pair is chosen with the probability.

$$P_{ij} = q(1-\frac{1}{2}\delta_{ij})c_i c_j \lambda_{ij} \tag{111}$$

In Eq. (111), $c_i$ is the relative concentration of $i$-th component, and $q$ is a normalized constant to ensure $\sum_{i,j} P_{ij} = 1$. Taking two-component plasma as an example, the probability of electron-electron, electron-ion and ion-ion collision is as follows.

$$P_{ee} = \frac{\lambda_{ee}}{\lambda_{ee}+\lambda_{ii}+2\lambda_{ei}}, P_{ei} = \frac{2\lambda_{ei}}{\lambda_{ee}+\lambda_{ii}+2\lambda_{ei}}, P_{ii} = \frac{\lambda_{ii}}{\lambda_{ee}+\lambda_{ii}+2\lambda_{ei}} \tag{112}$$

The three auxiliary parameters $\lambda_{ee}, \lambda_{ei}, \lambda_{ii}$ can be assumed to be unity in its simplest form. In order to speed up the simulation by increasing the time-step, Bobylev and Potapenko [56] recommend the optimal choice of the auxiliary parameters as follows

$$\lambda_{ee}=1, \lambda_{ei}=\lambda_{ie}=1/4, \lambda_{ii}=\sqrt{m_e/m_i} \tag{113}$$

Secondly, two particles are randomly chosen from the $i$-th and $j$-th component. The scattering angle is determined by the ratio of thermal velocity to relative velocity, instead of randomly sampled.

$$\cos\chi = \begin{cases} 1-2\varepsilon, 0<\varepsilon<1; \\ -1, \varepsilon \geq 1. \end{cases} \varepsilon = \frac{1}{\lambda_{ij}}\left(\frac{m_e q_i q_j}{m_{ij}e^2}\right)^2\left(\frac{v_{th}}{u}\right)^3 \tilde{\varepsilon} \tag{114}$$

### iv. MCC-like effective cross section method

Because of the $(1-\cos\chi)^2$ in the denominator of the Rutherford's DCS, the ICS is infinite. Therefore, the MCC-like method for Coulomb collision requires an effective cross section by cutoff of maximum impact parameter. It is natural that the cutoff of the maximum impact parameter $b_{max}$ by Debye shielding distance in plasma. Now the impact parameter is randomly sampled as

$$b = b_{max}\sqrt{r} = \lambda_{De}\sqrt{r} \tag{115}$$

Comparing Eq. (115) with that of Eq. (83),

$$\frac{b_0}{2\tan\frac{\chi}{2}} = \lambda_{De}\sqrt{r} \tag{116}$$

The scattering angle is randomly sampled as

$$\cos\chi = 1 - \frac{2}{1+\Lambda^2 r} \tag{117}$$

Eq. (117) is seldom used in sampling of scattering angle in Coulomb collision. The reason maybe that the ICS $\sigma_t = \pi\lambda_D^2$ scales with $\Lambda^2$, whose value is rather arbitrary. However, the momentum transfer cross section $\sigma_D = \pi b_0^2 \ln\Lambda$ only scales with $\ln\Lambda$, and the arbitrary choice of Coulomb logarithm has only a minor influence. It is not a bad idea that the effective cross section in MCC-like method scales with the momentum transfer cross section.

To account for the electron-electron collision effect on electron swarms in partially ionized gases, particle-mesh approaches to include electron-electron collision with isotropic scattering model were proposed in MC modeling [57, 58]. Their choice of effective cross section is exactly the momentum transfer cross section (note that their definition of $b_0$ is different from this paper). A similar effective-cross-section method is also proposed in our group [59] by considering the relation of impact parameter to average deflection angle Eq. (95)

$$\sigma_t = \pi b_{\max}^2 = \frac{2}{\langle\chi^2\rangle}\pi b_0^2 \ln\Lambda = k\sigma_D, k = \frac{2}{\langle\chi^2\rangle} \tag{118}$$

The probability of Coulomb collision for each particle pair is calculated as following

$$P_{coll} = 1 - \exp(-n\sigma_t v_r \Delta t) \tag{119}$$

The probability is compared to a random number uniformly distributed between [0, 1] to decide whether the collision occurs. The MCC-like method speeds up the Coulomb scattering by choosing a portion of colliding pairs instead of the whole. If the collision occurs, isotropic scattering model Eq. (70) is used to randomly sample the polar angle. Through massive MCC test on electron-electron and electron-ion collision, it is found that when $\langle \chi^2 \rangle = 0.5$, effective-cross-section method yields the accurate relaxation behavior with analytical example, and the point-wise error is comparable to that of TA's and Nanbu's method. More test should be performed for effective-cross-section method such as the modeling on Spitzer conductivity of ionized plasma.

### c) Ion-neutral collision

#### i. Polarization scattering with inverse-power potential

For ions with extremely high energy, the ion-neutral collision reduces to the aforementioned Rutherford scattering. For ions with common energies in LTP, polarization scattering of short range is the dominant process. The polarized potential can also be viewed as a special case of inverse-power potential

$$C = -\frac{q_1 q_2 \alpha_p}{8\pi\varepsilon_0}, n = 4 \tag{120}$$

Inserting the potential into Eq. (12), it is found that $r_{\min}$ is meaningful at a cutoff impact parameter

$$b_L = \left(\frac{\alpha_p q_1 q_2}{2\pi\varepsilon_0 E_r}\right)^{1/4} \tag{121}$$

The corresponding cross section is called Langevin cross section or capture cross section

$$\sigma_L = \pi b_L^2 = \sqrt{\frac{\pi \alpha_p q_1 q_2}{\varepsilon_0 m_r}} \frac{1}{v_r} \tag{122}$$

The cross section is representative of Maxwell's gas model $\sigma \propto v_r^{-1}$, whose collision frequency and probability is independent of impact velocity.

$$v_{coll} = n\sqrt{\frac{\pi \alpha_p q_1 q_2}{\varepsilon_0 m_r}} \Delta t \tag{123}$$

Resonant charge exchange between ions with their parent atoms is also an elastic scattering, and is not distinguishable from momentum exchange process in usual experiment conditions. Nanbu and Kitatani [60] developed the Langevin model in MC collision modeling of ion-neutral scattering. The non-dimensional impact parameter is defined as

$$\beta = b/b_L \tag{124}$$

The sampling process relies on two parameters: the maximum non-dimensional impact parameter $\beta_\infty$ and the limiting impact parameter below which charge-exchange (CX) is assumed possible $\beta_{CX}$. Firstly, the non-dimensional parameter is randomly sampled

$$\beta = \beta_\infty \sqrt{r} \tag{125}$$

If $\beta \leq 1$, the ion is "captured", and isotropic scattering model is assumed with charge-exchange probability $P_{CX}=1/2$; else if $1 < \beta < \beta_{CX}$, the charge-exchange probability is also 1/2; else if $\beta > \beta_{CX}$, no charge-exchange occurs; else $\beta > 3$, the ion trajectory is unchanged. $\beta_{CX}$ is chosen in such a way that the drift velocity resulting from this charge-exchange model agrees with the experimental data, and the exact value is suggested for rare gases colliding with parent atoms. The scattering angle is calculated according to Eq. (13) using the non-dimensional impact parameter if $\beta > 1$

$$\chi(\beta) = \pi - \frac{2\sqrt{2}\beta}{\epsilon_1} K(\zeta), \zeta = \epsilon_0/\epsilon_1,$$

$$\epsilon_0 = [\beta^2 - (\beta^4 - 1)^{1/2}]^{1/2}, \epsilon_1 = [\beta^2 + (\beta^4 - 1)^{1/2}]^{1/2} \tag{126}$$

In Eq. (126), $K(\zeta)$ is the complete elliptic integral of the first kind

$$K(\zeta) = \int_0^{\pi/2} (1 - \zeta^2 \sin^2\theta)^{-1/2} d\theta \tag{127}$$

### ii. Isotropic elastic scattering and backward charge exchange

Phelps suggested another angular scattering model for modeling ion-neutral collision [61]. The DCS is fitted by an isotropic part for pure elastic scattering and another back-scattering part for charge exchange

$$q(E, \chi) \approx \frac{1}{4\pi} \sigma_{iso} + \frac{\delta(\pi - \chi)}{2\pi \sin \chi} \sigma_{back} \tag{128}$$

The momentum transfer cross section is derived as

$$\sigma_D = \sigma_{iso} + 2\sigma_{back} \tag{129}$$

At high energies, the fitted back-scattering part $\sigma_{back}$ agrees well with measured charge-exchange cross sections [62], and the empirical formula by Sakabe and Izawa [63] can be used. At low energies, the isotropic part $\sigma_{iso}$ is well described by the Langevin cross section. At intermediate energies, the back-scattering part $\sigma_{back}$ is adjusted so that the momentum transfer cross section is consistent with that derived from measurements of ion mobility. Meanwhile, the smooth transition of the diffusion cross section to isotropic scattering cross section is assumed.

Scattering angle is sampled with the null-collision technique [64] when the DCS is decomposed into two parts. For the isotropic part, the scattering angle is sampled according to Eq. (70), and for the back-scattering part, the scattering angle is π.

### iii. Method consisting of both elastic and charge exchange collisions

A more reasonable approach to describe ion-neural scattering of identical particles is to directly use the DCS instead of decomposing it into an elastic and charge-exchange part. Measurements in real gases imply that the DCS is dominant at angles either close to π (which may be called charge exchange by convention) or close to zero degrees [65, 66]. The semi-classical approach gives the following expression of DCS according to classical theory Eq. (64)

$$q(E,\chi) = (1-P_{CX})\frac{d\sigma(\chi)}{d\Omega} + P_{CX}\frac{d\sigma(\pi-\chi)}{d\Omega}, \qquad (130)$$

The charge-exchange probability is

$$P_{CX} = \frac{1}{q(E,\chi)}\frac{d\sigma(\pi-\chi)}{d\Omega} \qquad (131)$$

For impact parameters of interest in LTP, $P_{CX}$ oscillates quickly between 0 and 1 with an average of 0.5, which leads to the common fitting formula of DCS as

$$q(E,\chi) = \frac{1}{2}\frac{d\sigma(\chi)}{d\Omega} + \frac{1}{2}\frac{d\sigma(\pi-\chi)}{d\Omega} \qquad (132)$$

Vahedi and Surendra [67] proposed a symmetric expression for the DCS

$$q(E_r,\chi) = \frac{A}{(1-\cos\chi+a)(1+\cos\chi+a)} \qquad (133)$$

In Eq. (133), both parameters $A$ and $a$ are a function of incident energy: $A$ is a normalization constant and $a$ is a small parameter to ensure the ICS is finite. The scattering angle is randomly sampled as

$$\cos \chi = (1+a)\frac{1-[a/(a+2)]^{1-2r}}{1+[a/(a+2)]^{1-2r}} \qquad (134)$$

However, the DCS in Eq. (133) does not conform to the small-angle approximation with inverse-power polarized potential as shown in Eq. (21-22)

$$\frac{d\sigma(\chi)}{d\Omega} \propto \frac{1}{\chi^{1.5}\sin\chi} \qquad (135)$$

Therefore, Wang et al. [20] suggested that small-angle scattering can be approximated by

$$\frac{d\sigma(\chi)}{d\Omega} \simeq \frac{A}{(1-\cos\chi+a)^{1.25}}, \qquad (136)$$

and the whole DCS is written in an asymmetric form

$$q(E_r,\chi) = \frac{A}{(1-\cos\chi+a)^{1.25}} + \frac{A}{(1+\cos\chi+b)^{1.25}} \qquad (137)$$

In Eq. (137), $b$ is another parameter with the similar role as that of $a$. Integrating the DCS, the total, diffusive, and viscosity cross sections are obtained, the fitting of which gives the three parameters $A$, $a$, and $b$. The randomly sampling of scattering angle can be performed by null-collision technique because the DCS is decomposed into two parts, namely

$$\cos\chi_1 = 1 + a - \{a^{-1/4} - r_1[a^{-1/4} - (2+a)^{-1/4}]\}^{-4} \qquad (138a)$$

$$\cos\chi_2 = -(1+b) + \{(2+b)^{-1/4} + r_2[b^{-1/4} - (2+b)^{-1/4}]\}^{-4} \qquad (138b)$$

### iv. Universal model from screened Coulomb potential

In general, the ion-neutral interacting potential is written as screened Coulomb potential.

$$V(r) = \frac{Z_1 Z_2 e^2}{4\pi\varepsilon_0 r}\chi(r) \qquad (139)$$

In Eq. (139), $\chi(r)$ is the screening factor, of which different forms have been proposed by Bohr, Thomas-Fermi, Moliere, Lenz-Jensen, and Ziegler-Biersack-Littmark (ZBL), respectively [68]. The ZBL screening factor reads as [69]

$$\chi(r) = \sum_{i=1}^{4} c_i \exp\left(-d_i \frac{r}{a_{ZBL}}\right) \qquad (140)$$

In Eq. (140), $c_i$ and $d_i$ are fitting parameters and $a_{ZBL}$ is the screening length given by

$$a_{ZBL} = 0.8854 \frac{a_0}{Z_1^{0.23}+Z_2^{0.23}} \tag{141}$$

In principle, the DCS of ion-neutral collision can be numerically obtained by inserting screened Coulomb potential Eq. (140) into Eq. (13) to derive the relation between impact parameter and scattering angle. However, the energy-transfer DCS with ZBL screening factor is defined in reduced notation as

$$d\sigma = \pi a_{ZBL}^2 \frac{f(x)}{2x^3} dx^2 \tag{142}$$

In Eq. (142), $x$ is the non-dimensional energy transferred during one collision event

$$x = \bar{E}\sin\frac{\chi}{2}, \bar{E} = \frac{4\pi\epsilon_0 a_{ZBL}}{Z_1 Z_2 e^2} \frac{m_2}{m_1+m_2} E_r \tag{143}$$

The function $f(x)$ can be obtained by inverting the reduced nuclear stopping power, of which analytical form is available

$$f(x) = \frac{d}{dx}(x S_n), S_n(x) = \frac{\ln(1+ax)}{2(x+bx^c+dx^{1/2})} \tag{144}$$

The four fitting coefficients ($a$, $b$, $c$, $d$) for specific ion-atom binary collision can be found in files of SRIM (The Stopping and Range of Ions in Matter). However, for the case of ZBL universal potential, the fitting coefficients are constants irrespective of the colliding pairs $a$=1.1383, $b$=0.01321, $c$=0.21226, $d$=0.19593. It was shown that the universal fitting coefficients yielded DCS with reasonable accuracy for most cases [70]. The sampling of scattering angle in ion-neutral collision with ZBL universal potential is not straightforward [71], because angle-dependent DCS is not invert-able.

$$\frac{d\sigma}{d\Omega} = \frac{d\sigma}{dx}\frac{dx}{d\Omega} \tag{145}$$

However, a look-up table can be generated on demand via analytical formulation.

### d) Electron-neutral collision

#### i. Elastic collision

##### 1. Electron collision with atom and non-polar molecular

Calculations of the DCS for electron-atom collision require making use of the quantum mechanical approach. FBA is commonly used to derive the angular scattering model. Let us first consider a simple screened Coulomb potential, namely the hydrogen-like Yukawa potential with screening length as Bohr radius.

$$V(r) = ZE_H \frac{a_0}{r} \exp(-\frac{r}{a_0})$$

(146)

Inserting Eq. (146) into Eq. (38), the normalized DCS is derived as [72]

$$I(\varepsilon, \chi) = \frac{1}{4\pi} \frac{1+8\varepsilon}{(1+4\varepsilon-4\varepsilon\cos\chi)^2},$$

(147)

In Eq. (147), the non-dimensional energy $\varepsilon$ is the incident electron energy scaled by $E_H$ one hartree in atomic unit $\varepsilon = E_r / E_H$. Eq. (147) represents the DCS of so-called Forward-Screened-Rutherford (FSR) type, upon which further improvements are built and will be discussed below.

For example, another commonly used FSR-type DCS is written as follows [73, 74]

$$I(\varepsilon, \chi) = \frac{1}{\pi} \frac{\eta(\eta+1)}{(2\eta+1-\cos\chi)^2}$$

(148)

Different from the non-dimensional energy $\varepsilon$ in Eq. (147), the screening factor $\eta$ in Eq. (148) depends not only on incident energy but also on the atomic charge $\eta = 1.89 Z^{2/3} / E[\text{eV}]$ as indicated by Mott [75].

However, both parameters $\varepsilon$ and $\eta$ are positive, resulting in forward-peaked DCS, which is generally not true for real gases. Phelps [76] used the FSR for elastic DCS but with an algebraic screening factor $\beta$

$$I(\varepsilon, \chi) = \frac{1}{4\pi} \frac{4\beta(1-\beta)}{(1-(1-2\beta)\cos\chi)^2}.$$

(149)

The screening factor $\beta$ for N$_2$ is found as follows

$$\beta = \frac{0.6}{[1+\sqrt{\varepsilon/50}+(\varepsilon/50)^{1.01}]^{0.99}}, \varepsilon = E[\text{eV}]/1$$

(150)

A fitting parameter $\xi(\varepsilon)$ is introduced by Okhrimovskyy *et al.* [77] in a general Screened-Rutherford (SR) type for normalized DCS

$$I(\varepsilon, \chi) = \frac{1}{4\pi} \frac{1-\xi^2(\varepsilon)}{(1-\xi(\varepsilon)\cos\chi)^2}$$

(151)

The DCS in Eq. (151) is quite universal and is readily applied to non-polar molecule. The fitting parameter in MC collision modeling is reported for He, and N$_2$, to name only a few. The value of $\xi(\varepsilon)$ is obtained by fitting the ratio between the measured ICS and diffusive cross section

$$\frac{\sigma_D}{\sigma_t} = \frac{1-\xi}{2\xi^2}((1+\xi)\ln\frac{1+\xi}{1-\xi} - 2\xi) \tag{152}$$

$\xi(\varepsilon)$ is in the range (-1, 1): a positive value implies FSR-type DCS; a negative value implies Backward-Screened-Rutherford (BSR) DCS; and a value of zero implies isotropic DCS. The randomly sampling of the scattering angle depends on the fitting parameter

$$\cos\chi = 1 - \frac{2r(1-\xi)}{1+\xi(1-2r)} \tag{153}$$

The DCS in Eqs. (147-149) can be viewed as a special case of the general SR-type DCS in Eq. (151), if the fitting parameter is set in the following way.

$$\xi(\varepsilon) = \frac{4\varepsilon}{1+4\varepsilon} \tag{154a}$$

$$\xi(\varepsilon) = \frac{1}{1+2\eta} \tag{154b}$$

$$\xi(\varepsilon) = 1 - 2\beta \tag{154c}$$

Recently, Janssen et al. [17] proposed that the DCS is a weighted combination of both isotropic scattering and FSR

$$I(\varepsilon,\chi) = \frac{C}{4\pi} + \frac{1-C}{\pi}\frac{\eta(\eta+1)}{(2\eta+1-\cos\chi)^2} \tag{155}$$

In Eq. (155), the weighting factor $C$ and the screening factor $\eta$ can be obtained by fitting the measured total, diffusive and viscosity cross section.

$$\frac{\sigma_D}{\sigma_t} = C + (1-C)[2\eta(\eta+1)\ln(1+1/\eta) - 2\eta] \tag{156a}$$

$$\frac{\sigma_v}{\sigma_t} = \frac{2}{3}C + (1-C)4\eta(\eta+1)[(2\eta+1)\ln(1+1/\eta) - 2] \tag{156b}$$

A complicated sampling formula for scattering angle is derived.

$$\cos\chi = \frac{1-r+\eta-\sqrt{(\eta+1-C-r)^2+4C\eta r}}{C} \tag{157}$$

Murphy [78] proposed another DCS which is a weighted combination of both FSR and BSR

$$I(E,\chi) = C\frac{\eta_F(\eta_F+1)}{\pi(2\eta_F+1-\cos\chi)^2} + (1-C)\frac{\eta_B(\eta_B+1)}{\pi(2\eta_B+1+\cos\chi)^2} \tag{158}$$

The weighting factor $C$ and the screening factor for forward and backward scattering can be obtained in a similar way, and fitting formula is reported for He, $N_2$ and $O_2$, to name only a few. Park et al. [26] proposed a rather complicated formula for randomly sampling the scattering angle. However, with the aid of null-collision technique, the sampling process is easily performed for either part of FSR or BSR (the value of $C$ is in the range between 0 and 1 for physically sound model)

$$\cos\chi_1 = \frac{\eta_F+1-r_1(1+2\eta_F)}{\eta_F+1-r_1}, 0 < r \leq C \tag{159a}$$

$$\cos\chi_2 = \frac{r_2(1+2\eta_B)-\eta_B}{\eta_B+r_2}, C < r < 1 \tag{159b}$$

The normalized DCS is added by another term in fitting experimental data [79], considering the evidence that at high energies the DCS follows a negative exponential trend with $\chi$ for small angles

$$I(E,\chi) = C_1\frac{1+\alpha^2}{2\pi(1+e^{-\alpha\pi})}e^{-\alpha\chi} + C_2\frac{\eta_F(\eta_F+1)}{\pi(2\eta_F+1-\cos\chi)^2} + (1-C_1-C_2)\frac{\eta_B(\eta_B+1)}{\pi(2\eta_B+1+\cos\chi)^2} \tag{160}$$

Directly sampling of the scattering angle according to Eq. (160) is so complicated that the null-collision technique is suggested. The scattering angle for the last two terms in the RHS is already shown in Eq. (159), and here only the first term is shown.

$$1-e^{-\alpha\chi}(\alpha\sin\chi+\cos\chi) = r(1+e^{-\alpha\pi}) \tag{161}$$

Regarding the first term in Eq. (160), Boeuf and Marode [12] proposed a similar exponential decay DCS for high-energy electron collision with He

$$q(E,\chi) = Ce^{-aE^{1/2}\sin(\chi/2)} \tag{162}$$

For medium-energy electron collision with Cl, Nanbu [80] proposed another DCS by combining the exponential decay term with a symmetric form for large-angle scattering

$$I(E,\chi) = Ae^{-\alpha\chi} + Be^{-\beta(\pi-\chi)} \tag{163}$$

A truncated series of Legendre polynomials (LP) in cosχ is added to the FSR to fit the experimental DCS [23], in order to characterize the large angle or back scattering

$$I(\varepsilon,\chi) = \frac{C}{\pi} \frac{\eta(\eta+1)}{(2\eta+1-\cos\chi)^2} + \sum_{l=0}^{L} A_l P_l(\cos\chi) \qquad (164)$$

Randomly sampling of the scattering angle can be performed by the aid of null-collision technique: the scattering angle by first term is already shown in Eq. (159a), and the scattering angle by the second term LP is shown in the Appendix b.

Moss *et al.* [81] modified the FSR by moving the screening factor $\eta$ out of the bracket in the denominator at electron energies below 500 eV for $N_2$ and 1 keV for $O_2$

$$I(\varepsilon,\chi) = \frac{1}{2\pi \arctan(2/\eta)} \frac{\eta}{(1-\cos\chi)^2 + \eta^2}, \eta = E_r[\text{eV}]/4 \qquad (165)$$

The scattering angle is randomly sampled as

$$\cos\chi = 1 - \eta \tan(r \arctan\frac{2}{\eta}) \qquad (166)$$

The electron-neutral interacting potential is not always like the Yukawa's potential shown in Eq. (146). For example, electron-$Cl_2$ interacting potential is proposed to decay with distance as follows

$$V(r) = V_0 \exp(-\frac{r^2}{a^2}) \qquad (167)$$

Inserting Eq. (167) into the FBA, the normalized DCS follows the Gaussian distribution (similar to Nanbu's method in Coulomb collision) [80]

$$I(E,\chi) = \frac{\alpha}{4\pi(1-\exp(-\alpha))} \exp(-\alpha \sin^2\frac{\chi}{2}), \alpha = 2(ka)^2 \qquad (168)$$

The scattering angle is randomly sampled as

$$\cos\chi = 1 + \frac{2}{\alpha} \ln[r(1-e^{-\alpha}) + e^{-\alpha}] \qquad (169)$$

There is also a simple method to represent the anisotropic effect of DCS by decomposing it into two isotropic parts: one for forward scattering and the other for back scattering [82]. Random sampling of the scattering angle is performed into two steps: the first step to determine whether forward or backward scattering occurs; and the second step to determine the scattering angle from Eq. (70).

## 2. Electron collision with polar molecules

Electron collisions with polar molecules are characterized by a dipole moment [83], and therefore a SR-like DCS from Born approximation may be not valid. To not lose generality, the interacting potential is written as

$$V(\mathbf{r}; \mathbf{R}) = \sum_\lambda \frac{eM_\lambda}{r^{\lambda+1}} P_\lambda(\hat{\mathbf{r}} \cdot \hat{\mathbf{R}}) \tag{170}$$

In Eq. (170), $M_\lambda$ represents the permanent electric multi-pole moment, with $\lambda = 1, 2$ stands for dipole and quadrupole moment, respectively. $\hat{\mathbf{r}}$ and $\hat{\mathbf{R}}$ are the unit vector of electron position and molecular position, respectively. Inserting Eq. (170) into the FBA, the scattering amplitude is obtained as

$$f(\chi, E) = -\frac{i^\lambda \sqrt{\pi} e M_\lambda m_r}{\hbar^2 2^{\lambda-1} \Gamma(\lambda+1/2) K^{2-\lambda}} P_\lambda(\hat{\mathbf{R}}) \tag{171}$$

$\lambda = 1$, the DCS for dipole moment is

$$q(\chi, E) = [\frac{2eM_1 m_r}{\hbar^2 K} P_1(\hat{\mathbf{R}})]^2 \tag{172}$$

We are interested in the angular scattering model, instead of the absolute value of the DCS

$$I(\chi, E) \propto \frac{1}{K^2} = \frac{1}{2k^2(1-\cos\chi)} \tag{173}$$

The ICS is infinite unless the scattering angle is cutoff.

### ii. Inelastic collision

The dipole moment interaction dominates the rotational excitation of electron collision with molecules at low energies. In adiabatic approximation [84], the rotational excitation DCS is related to the elastic scattering amplitude in Eq. (171) as

$$q_{rot}(\theta) = \frac{k_{sca}}{k} |\int d^3\mathbf{R}\, Y_{jm}(\hat{\mathbf{R}}) f(\chi, \varphi) Y^*_{j'm'}(\hat{\mathbf{R}})|^2 \tag{174}$$

Inserting the scattering amplitude for dipole moment Eq. (171) into Eq. (174),

$$q_{rot}(\theta) = \frac{4\pi k_{sca}}{3k}\left(\frac{2eM_1m_r}{\hbar^2 K}\right)^2 \left|\int d^3\mathbf{R}\, Y_{jm}(\widehat{\mathbf{R}})Y_{10}(\widehat{\mathbf{R}})Y^*_{j'm'}(\widehat{\mathbf{R}})\right|^2 \quad (175)$$

The integration in the RHS can be performed with the aid of Clebsch-Gordan coefficient, whose properties give the selecting rule of rotational transition

$$j' = j \pm 1, m' = m \quad (176)$$

Averaging over the quantum number *m*, the DCS for rotational excitation $j' = j + 1$ is finally obtained as

$$q_{rot}^{j\to j+1}(\chi) = \frac{k_{sca}}{3k}\left(\frac{2eM_1m_r}{\hbar^2 K}\right)^2 \frac{j+1}{2j+1}, \quad (177)$$

The angular scattering model is written as

$$I(\varepsilon,\chi) = \frac{1}{2\pi}\frac{\beta}{(1+\beta^2-2\beta\cos\chi)\ln\left(\frac{1+\beta}{1-\beta}\right)}, \beta = \frac{k_{sca}}{k} = \sqrt{1-\frac{E_{th}}{E}} \quad (178)$$

The scattering angle is randomly sampled as

$$\cos\chi = \frac{1+\beta^2-(1-\beta)^2[(1+\beta)/(1-\beta)]^{2r}}{2\beta} \quad (179)$$

In the open-source Lisbon Kinetics Boltzmann solver LoKI-B, other non-dimensional parameters *x* and *ζ* are used [24].

$$x = \frac{E}{E_{th}} = \frac{1}{1-\beta^2}, \xi = \frac{E_{th}}{(\sqrt{E}+\sqrt{E+E_{th}})^2} = \frac{1-\beta}{1+\beta} \quad (180)$$

Now the scattering angle is randomly sampled by

$$\cos\chi = 1 + \frac{2\xi^2}{1-\xi^2}(1-\xi^{-2r}) \quad (181)$$

For rotational excitation, the incident energy is usually much higher than the collision threshold $\beta \to 1, \xi \to 0$, The scattering angle is randomly sampled in a simple way

$$\cos\chi = 1 - 2\xi^{-2(r-1)} \quad (182)$$

The DCS for a quad-rupole moment can be derived in a similar way and is found to be independent of the transferred momentum *K*. **The DCS for rotational excitation can be readily extended to vibrational excitation if FBA is used** [85].

There is, however, another way to derive Eq. (178) in Bethe approximation.

$$q^{Bethe} = \frac{2m_r z^2 e^4}{\hbar^2 E_n} \frac{k_{sca}}{k} \frac{1}{K^2} f_n(K) \tag{183}$$

For small angle scattering, the momentum change approaches zero $K \to 0$, the OOS is used in Eq. (183) instead of the GOS, and now

$$q^{Bethe} \propto \frac{1}{K^2} \tag{184}$$

The normalized DCS is the same as in Eq. (178). Note that Eq. (184) is derived by **Born-Bethe approximation at small scattering angle**.

A widely used empirical DCS but with incorrect non-dimensional energy was initially proposed by Surendra *et al* [86] for elastic scattering

$$I(\varepsilon,\chi) = \frac{1}{4\pi} \frac{\varepsilon}{(1+\varepsilon \sin^2(\chi/2))\ln(1+\varepsilon)}, \varepsilon = E/1\text{eV} \tag{185}$$

If the non-dimensional energy is scaled as

$$\varepsilon = \frac{4\beta}{(1-\beta)^2}, \beta = \frac{k_{sca}}{k} = \sqrt{1-\frac{E_{th}}{E}} \tag{186}$$

Eq. (185) reduces to the more general and correct form Eq. (178). The scattering angle is randomly sampled according to Eq. (185) in a simple form as

$$\cos\chi = \frac{2+\varepsilon - 2(1+\varepsilon)^r}{\varepsilon}, \varepsilon = E/1\text{eV} \tag{187}$$

There are also some empirical formulas to represent DCS in inelastic scattering. The first one is a delta-function of scattering angle, by which the projectile does not change its direction after inelastic scattering

$$I(\varepsilon,\chi) = \frac{1}{2\pi} \frac{\delta(\chi)}{\sin\chi}, \tag{188}$$

Another anisotropic DCS was proposed by Kushner [87]

$$f(\varepsilon,\chi) = \frac{n(\varepsilon)+2}{8\pi} \cos^n \frac{\chi}{2} \tag{189}$$

The scattering angle is randomly sampled as

$$\cos\chi = 2(1-r)^{\frac{2}{n+2}} - 1 \tag{190}$$

### iii. Ionization

#### 1. The complete DCS

The MC collision modeling on ionization is more complicated: energy partition model is as important as the angular scattering models [88-90]. However, it is difficult to discern which out-going electron is the scattered projectile or the new born one in experiment. Usually, the electron with lower energy is called the ejected or secondary electron, and the other one is called scattered or primary electron. By FBA, the triply DCS (TDCS) is calculated

$$q^{(3)} = \frac{d\sigma}{d\Omega_s d\Omega_e dE_e} = \frac{1}{2\pi} \frac{k_e k_s}{k_i} \langle \Psi_f | V | \Psi_i \rangle \tag{191}$$

In Eq. (191), the subscript of i, e, and s stands for incident, ejected and scattered electron, respectively. Integrating the TDCS in the solid angle of scattered electron, the doubly DCS (DDCS) is derived

$$q^{(2)} = \int q^{(3)} d\Omega_s = \frac{d\sigma}{d\Omega_e dE_e} \tag{192}$$

Integrating the DDCS in the solid angle of scattered electron, the singly DCS (SDCS) is derived

$$q^{(1)} = \int q^{(2)} d\Omega_e = \frac{d\sigma}{dE_e} \tag{193}$$

The total ionization cross section (TICS) is obtained by integrating the SDCS from zero to half the available energy, namely incident energy minus the ionization energy $I_{ion}$.

$$\sigma_{ion} = \int_0^{E_{max}} q^{(1)}(E_i, E_e) dE_e, \quad E_{max} = \frac{E_i - I_{ion}}{2} \tag{194}$$

## 2. The energy partition model

In MC collision modeling on ionization, the first step is usually to determine how the energy is partitioned between the scattered and ejected electrons. There are deterministic models as well as probabilistic models. A simple deterministic model assumes equal energy partition between the scattered and ejected electrons. Another commonly used deterministic model is that the scattered electron takes away all the available energy, while the ejected electron is born with zero kinetic energy. Yoshida *et al* proposed another deterministic model for ejected electron energy [88]:

$$E_2 = \begin{cases} (E_i - I_{\text{ion}})/2, & E_i < I_{\text{ion}} + 2u \\ u, & E_i \geq I_{\text{ion}} + 2u \end{cases} \tag{195}$$

Eq. (195) means that below an energy threshold $I_{\text{ion}} + 2u$, equal energy partition is used; above the threshold, the energy of ejected electron is constant $u$, which is adjusted in MC collision modeling to fit the swarm coefficients.

A simple probabilistic model is random energy partition, which assumes that the energy of ejected electron is uniformly distributed between zero kinetic energy and the available energy [90].

$$E_2 = r(E_i - I_{\text{ion}})/2 \tag{196}$$

More complicated probabilistic models for energy partition usually rely on the SDCS.

$$r\sigma_{ion} = \int_0^E q^{(1)}(E_i, E_e) \, dE_e \tag{197}$$

Opal *et al* [91] fitted the experimental SDCS, and found the following empirical form

$$q^{(1)}(E_i, E_e) \propto \frac{1}{1 + [E_e/B(E_i)]^{2.1}} \tag{198}$$

In Eq. (198), $B$ is a parameter which is the order of ionization energy. In order to use Eq. (198) in MC collision modeling, the SDCS is slightly modified to a Cauchy-Lorentz distribution

$$q^{(1)}(E_i, E_e) = \frac{\sigma_{\text{TICS}}}{B \arctan(\frac{E_i - I_{ion}}{2B})} \frac{1}{1 + (E_e/B)^2} \tag{199}$$

Inserting Eq. (199) into Eq. (196), the energy partition is randomly sampled as

$$E_2 = B \tan\{r \arctan[(E_i - I_{ion})/2B]\} \tag{200}$$

**When the incident energy is slightly above the ionization threshold**, namely

$$(E_i - I_{ion})/2B \ll 1 \tag{201}$$

The sampling is reduced to random energy partition based on the approximation $\tan x \simeq x$.

Later, the general Cauchy distribution was proposed to fit the experimental SDCS with a parameter of energy shift $E_0$ [92, 93] (in some literature called the Breit-Wigner shape)

$$q^{(1)}(E_i, E_e) = \frac{\sigma_{TICS}}{B[\arctan(\frac{E_1}{B}) + \arctan(\frac{E_0}{B})]} \frac{1}{1 + (\frac{E_e - E_0}{B})^2}, E_1 = \frac{E_i - I_{ion}}{2} - E_0 \tag{202}$$

The energy partition is randomly sampled by

$$E_2 = E_0 + B \tan[r \arctan(E_1/B) + (r-1)\arctan(E_0/B)] \tag{203}$$

If $E_0$ equals to zero, Eq. (203) reduces to the commonly used Eq. (200).

The Binary-Encounter model is very attractive in MC collision modeling on ionization [94], due to their rather simple and analytical expressions both for the TICS and SDCS, and their applicability to a large set of elements and/or subshells. Let us first return to the Mott's DCS in Eq. (92) at high energy limit (the third term in the cosine function is approximated to be unity) and extend it to the case of electron-collision ionization of target atom with $N$ bound electrons

$$\frac{d\sigma^{Mott}}{dW} = \frac{4\pi N a_0^2 E_{Ryd}^2}{E}[\frac{1}{W^2} + \frac{1}{(E-W)^2} - \frac{1}{W(E-W)}] \tag{204}$$

The Mott's DCS is not consistent when applied to ionization directly. Modifications should be made by considering the following three issues: firstly, the SDCS should be symmetric under the exchange of the energies of the two electrons [95], namely $W + I_{ion}$ and $E - W$; secondly, to account for the attraction of the projectile electron by the nucleus, the SDCS should be multiplied by a so-called "acceleration factor" $E/(E+I_{ion}+U)$ in which $U$ is the average kinetic energy of bound electron; thirdly, the SDCS is expected to include an extra term proportional to $U$ in accordance with the classical binary-encounter approximation. The above modifications lead to a non-relativistic semi-classical SDCS with impulse approximation ($I$ stands for ionization threshold)

$$\frac{d\sigma^{BE}}{dW} = \frac{4\pi N a_0^2 E_{Ryd}^2}{E+I+U} \{\frac{1}{(W+I)^2} + \frac{1}{(E-W)^2} - \frac{1}{(W+I)(E-W)} + \frac{4U}{3}[\frac{1}{(W+I)^3} + \frac{1}{(E-W)^3}]\}$$

(205)

It is more convenient to write the DCS into dimensionless form

$$\frac{d\sigma^{BE}}{dw} = \frac{S}{t+1+u} \{\frac{1}{(w+1)^2} + \frac{1}{(t-w)^2} - \frac{1}{(w+1)(t-w)} + \frac{4u}{3}[\frac{1}{(w+1)^3} + \frac{1}{(t-w)^3}]\}$$

$$S = 4\pi N a_0^2 \frac{E_{Ryd}^2}{I_{ion}^2}, t = \frac{E}{I_{ion}}, w = \frac{W}{I_{ion}}, u = \frac{U}{I_{ion}}$$

(206)

Both the Mott and binary-encounter DCS can be recast into series

$$\frac{d\sigma^{BE}}{dw} = S\sum_{n=1}^{3} F_n(t)[\frac{1}{(w+1)^n} + \frac{1}{(t-w)^n}]$$

(207)

For Mott's SDCS, we have

$$F_1 = -\frac{F_2}{t+1}, F_2 = \frac{1}{t}, F_3 = 0$$

(208)

For binary-encounter SDCS, we have

$$F_1 = -\frac{F_2}{t+1}, F_2 = \frac{1}{t+u+1}, F_3 = \frac{4u}{3}F_2$$

(209)

However, the binary-encounter SDCS is not successful to reproduce experimental SDCS, and does not approach the high-energy asymptotic behavior given by Bethe [34] either. To circumvent this failure, the following binary-encounter-Bethe model was proposed [96]

$$F_1 = -\frac{F_2\phi}{t+1}, F_2 = \frac{2-Q}{t+u+1}, F_3 = \frac{Q\ln t}{t+u+1}$$

(210)

In Eq. (210), $\phi$ is the Vriens function [97]

$$\phi = \cos(\sqrt{\frac{E_{Ryd}}{I_{ion}(t+1)}} \ln t) \tag{211}$$

and $Q$ is given by the differential oscillator strength

$$Q = \frac{N_i}{N} = \frac{\int_0^\infty \frac{df}{dw} dw}{N} \tag{212}$$

The indefinite integral of the SDCS is derived as

$$H(w) = S\{F_1 \ln \frac{w+1}{t-w} + F_2(\frac{1}{t-w} - \frac{1}{w+1}) + \frac{F_3}{2}[\frac{1}{(t-w)^2} - \frac{1}{(w+1)^2}]\} \tag{213}$$

The TICS is obtained by integration in the range $[0, \frac{t-1}{2}]$

$$\sigma^{BEB} = H(\frac{t-1}{2}) - H(0) = S[F_1 \ln t + F_2(1-\frac{1}{t}) + \frac{F_3}{2}(1-\frac{1}{t^2})] \tag{214}$$

The energy partition is randomly sampled as

$$H(w) = (1-r)H(0) \tag{215}$$

If no information is available for the differential oscillator strength, Kim and Rudd [96] set $Q=1$. The Vriens function can also be approximated by its asymptotic value as $\phi \simeq 1$ if the ionization threshold is not smaller than the Rydberg constant. Inserting the above value into Eq. (214), we have

$$\sigma^{BEB} = \frac{S}{t+u+1}[-\frac{1}{t+1}\ln t + (1-\frac{1}{t}) + \frac{\ln t}{2}(1-\frac{1}{t^2})] \tag{216}$$

### 3. The angular scattering model

Once the energy of scattered and ejected electron is known, scattering angle is usually sampled by one of the formulas provided in the above sub-sections using either elastic or inelastic scattering. For example, Nanbu [80] assumed that in the COM frame, the post collision velocities of two electrons are isotropic. However, there are some angular scattering models uniquely applied to ionization collision.

Based on following assumptions: firstly, the incident, ejected and scattered electron velocities are co-planar; secondly, the scattered and ejected electron velocities are perpendicular, Boeuf and Marode [12] proposed a simple method to obtain the scattering angle

$$\cos \chi_s = (\frac{E_s}{E_i - I_{ion}})^{0.5}, \cos \chi_e = (\frac{E_e}{E_i - I_{ion}})^{0.5} \qquad (217)$$

Note that both scattering angles in Eq. (217) are in forward direction [0, $\pi/2$], and only determined by the scattered or ejected energy, instead of sampled by random numbers. In fact, the assumptions proposed by Boeuf and Marode [12] are not generally true in ionization collision and their assumptions alone cannot lead to the scattering angle in Eq. (217). Nevertheless, the simple angular scattering law has been adopted in many MC collision codes, including the introductory PIC code eduPIC [98], in which a constraint on azimuth angles is imposed.

$$\phi_s = 2\pi r, \phi_e = \phi_s + \pi \qquad (218)$$

A random scatter model for ionization is proposed recently [27]. A restriction is imposed on the model that the ejected electron's scattering direction is determined randomly over a forward hemisphere which is perpendicular to the incident electron vector. The velocity vector of scattered electron is determined from momentum conservation by ignoring the momentum of heavy particle. As a result, both scattered and ejected electron are forward scattered relative to the incident electron's unit vector during ionization events.

Nguyen *et al.* [90] did not sample the scattering angle χ between ejected electron and incident electron, but the angle Ψ between ejected electron and scattered electron. Based on Wannier's classical treatment, the PDF of Ψ is a Gaussian distribution

$$f(\Psi) = k \exp[-\frac{(\Psi - \pi)^2}{2\sigma^2}], \sigma = 3.5(E_{inc}[\text{eV}] - I_{ion}[\text{eV}])^{1/4} \qquad (219)$$

However, the model needs another constraint to solve the post-collision velocity. Nguyen *et al.* [90] imposed momentum conservation similar to that in random scatter model [27].

A note should be put here then: in energy partition model, the kinetic energy of heavy particle in energy conservation can be ignored, **but the momentum of heavy particle in momentum conservation during electron-impact ionization cannot be generally ignored** [80]!

Until very recently, the DDCS by FBA (for example Eq. (6) for ejected electron and Eq. (9) for scattered electron in Ref. [99]) is used in MC collision modeling on ionization to determine the scattering angle [100, 101]. The expressions of the DDCS (in atomic unit) for scattered electron was given for electron ionization of helium by FBA [102]

$$\frac{d\sigma}{d\Omega_s dE_s} = 2^{10} \frac{k_s k_e}{k_i} \frac{1}{K^2} \frac{[K^2 + \frac{1}{3}(1+k_e^2)]\exp(-\frac{2}{k_e}\arctan\frac{2k_e}{K^2 - k_e^2 + 1})}{[(K+k_e)^2 + 1]^3 [(K-k_e)^2 + 1]^3 (1-\exp\frac{-2\pi}{k_e})},$$
(220)

The DDCS for ejected electron is more complicated than Eq. (220) and is not listed here. Under common energies of incident electron, it is found that the shape of DDCS for hydrogen and helium is reasonably fitted by two Lorentz distributions following the binary-encounter-Bethe model [103].

$$q^{(2)}(w,t,\chi) = G_1[f_{BE}(w,t,\chi) + G_4 f_b(w,t,\chi)] \quad (221)$$

In Eq. (221), $f_{BE}$ and $f_b$ denote the binary encounter peak and back-scattering peak, respectively

$$f_{BE}(w,t,\chi) = \frac{1}{1+(\frac{\cos\chi - G_2}{G_3})^2}, f_b(w,t,\chi) = \frac{1}{1+(\frac{\cos\chi + 1}{G_5})^2} \quad (222)$$

The parameters $G_2 \sim G_5$ are obtained by fitting the experimental DDCS, and for hydrogen and helium

$$G_2 = (\frac{w+1}{t})^{1/2}, G_3 = 0.6(\frac{1-G_2^2}{w})^{1/2}, G_4 = 10\frac{(1-w/t)^3}{t(w+1)}, G_5 = 0.33. \quad (223)$$

The parameter $G_1$ is used to adjust the integrated SDCS.

$$G_1 = \frac{q^{(1)}/(2\pi)}{g_{BE}(w,t) + G_4 g_b(w,t)},$$

$$g_{BE} = G_3(\arctan\frac{1-G_2}{G_3} + \arctan\frac{1+G_2}{G_3}), g_b = G_5 \arctan\frac{2}{G_5} \qquad (224)$$

Because the DDCS is decomposed into two parts, sampling of the scattering angle of secondary electron in MC scheme is performed with null-collision technique.

$$\cos\chi_1 = G_2 + G_3 \tan[(1-r_1)\arctan\frac{1-G_2}{G_3} - r_1 \arctan\frac{1+G_2}{G_3}] \qquad (225a)$$

$$\cos\chi_2 = -1 + G_5 \tan[(1-r_2)\arctan\frac{2}{G_5}] \qquad (225b)$$

## V. Summary

In summary, a tutorial overview of angular scattering models in MC collision modeling on low temperature plasma is given. The classical scattering theory and the most used inverse-power law are firstly described. The quantum scattering theory is also introduced, with the commonly used Born-Bethe approximation, partial wave analysis, and effective range theory. State-of-the-art angular scattering models for electron-neutral, ion-neutral, neutral-neutral, and Coulomb collisions are given in detail, with an emphasis on how those models are derived from classical, semi-classical, and perturbation theories of quantum mechanics. The reviewed models are expected to be readily used in MC scheme, in which scattering angles can be sampled by easily-invertible formula. The energy partition between scattered and ejected electrons in electron-neutral ionization is also discussed, and the attractive model based on binary encounter Bethe theory is mentioned. The appropriate model of angular scattering for specific projectile-target combination is left to the user of MC collision code, with possible insight from this tutorial overview.

## VI. Appendix

a) **Sampling of scattering angle by impact parameter *b* or wave-vector change *K***

In classical theory, random sampling of the classical impact parameter $b$ is more convenient than Eq. (67),

$$\int_0^{b_\chi} b\,\mathrm{d}b = r \int_0^{b_{max}} b\,\mathrm{d}b \tag{A1}$$

It is written as

$$b_\chi = b_{max}\sqrt{r} \tag{A2}$$

Taking the VSS as an example, we have

$$b_{max} = d,\, b_\chi = d\cos^\alpha \frac{\chi}{2} \tag{A3}$$

Inserting Eq. (A3) into Eq. (A2), it is simple to determine the scattering angle

$$\cos^\alpha \frac{\chi}{2} = \sqrt{r} \tag{A4}$$

from which the randomly sampling of the scattering angle is derived as

$$\cos\chi = 2r^{1/\alpha} - 1 \tag{A5}$$

Eq. (A5) is equivalent to Eq. (81) if we replace $1-r$ as $r$.

In quantum theory, random sampling of the wave-vector change $K$ is more convenient.

$$\int_{K_{min}}^{K_\chi} \frac{\mathrm{d}\sigma}{\mathrm{d}K} K\,\mathrm{d}K = r \int_{K_{min}}^{K_{max}} \frac{\mathrm{d}\sigma}{\mathrm{d}K} K\,\mathrm{d}K,\, K_{min} = k - k_{sca},\, K_{max} = k + k_{sca} \tag{A6}$$

Taking the rotational excitation as an example,

$$\frac{\mathrm{d}\sigma}{\mathrm{d}K} \propto \frac{1}{K^2} \tag{A7}$$

Inserting Eq. (A7) into Eq. (A6), we have

$$\ln \frac{K_\chi}{K_{min}} = r \ln \frac{K_{max}}{K_{min}} \qquad (A8)$$

then scattering angle is derived by the sampled $K$

$$K_\chi^2 = k^2 + k_{sca}^2 - 2kk_{sca} \cos \chi \qquad (A9)$$

**b) Sampling of scattering angle for combinations of Legendre polynomial**

In partial wave analysis, the DCS is written as a function of Legendre polynomial. Using Rodrigues's formula, the Legendre polynomial is transformed into power series

$$P_l(x) = \frac{1}{2^l l!} \frac{d^l}{dx^l}(x^2 - 1)^l, x = \cos\chi \qquad (A10)$$

In general, the normalized DCS can be fitted as a power series of $\cos\chi$

$$I = \frac{1}{4\pi\alpha_{even}} \sum_{n=0}^{\infty} A_n \cos^n\chi \qquad (A11)$$

For example, Belenguer and Pitchford [14] proposed the differential elastic scattering cross section

$$I = I_0 + I_1 \cos \chi + I_2 \cos^2 \chi \qquad (A12)$$

The normalization constant $\alpha_{even}$ is given by

$$\alpha_{even} = \sum_{m=0}^{\infty} \frac{A_{2m}}{2m+1} \qquad (A13)$$

The parameter $\alpha_{odd}$ is similar to $\alpha_{even}$

$$\alpha_{odd} = \sum_{m=0}^{\infty} \frac{A_{2m+1}}{2m+2} \qquad (A14)$$

The scattering angle is randomly sampled as

$$\frac{1}{2}\left(1 + \frac{\alpha_{odd}}{\alpha_{even}} - \frac{1}{\alpha_{even}} \sum_{n=0}^{\infty} \frac{A_n}{n+1} \cos^n\chi\right) = r \qquad (A15)$$

**c) Sampling of scattering angle for weighted particle scattering**

For binary particle collision with different weights, the sampling of scattering angle is generally the same as that for equally weighted collision pair. It is not reasonable to use different formula to sample the scattering angle with non-equal weighted pair [104]. The details of weighted particle collision are well documented in the literature [105-108]. The post-scattering velocity of large-weighted particle is updated with a probability or only portion of the large-weighted particle undergoes collision and then merges.